\documentclass[final,3p,times,twocolumn,compress]{elsarticle}

\usepackage{amssymb}
\usepackage{amsmath}

\usepackage{url}
\usepackage{xcolor}
\definecolor{newcolor}{rgb}{.8,.349,.1}
\usepackage{hyperref}
\usepackage[switch,pagewise]{lineno} 
\usepackage{amsmath}
\usepackage{tabularx,multirow,booktabs,blindtext}
\usepackage{array}
\usepackage{soul}
\usepackage{enumitem}
\usepackage{comment}
\usepackage{balance}
\usepackage{natbib}

\begin{document}

\begin{frontmatter}

\title{Shear-layer effects on the dynamics of unsteady premixed laminar counterflow flames}

\author{Jose G Rivera Lizarralde} 

\author{Aditya Potnis\fnref{label2}} 
\fntext[label2]{present address: School of Engineering, Institute for Energy Systems, University of Edinburgh, Edinburgh, EH9 3FB, UK}

\author{Abhishek Saha\corref{cor1}} 
\cortext[cor1]{Corresponding author. Email: asaha@ucsd.edu}

\affiliation{Department of Mechanical and Aerospace Engineering, University of California San Diego, La Jolla, CA 92093, USA}


\begin{abstract}
The influence of flow non-uniformity and unsteadiness on premixed flames is of considerable interest due to its direct relevance to practical combustion systems. The steady counterflow flame has long served as a canonical configuration for investigating flame dynamics under controlled, spatially non-uniform conditions. A commonly studied variation, referred to as the unsteady counterflow, introduces a controlled temporal perturbation to the otherwise steady flow from the nozzles, thereby enabling the systematic examination of the coupled effects of unsteadiness and non-uniformity. Prior investigations have focused on flame dynamics along the line of symmetry, where the reduced dimensionality of the problem facilitates analysis. In the present study, we extend this perspective by experimentally examining flame behavior at off-center locations, where multi-dimensional effects of non-uniformity and unsteadiness are more pronounced. Results reveal markedly different dynamics away from the centerline, characterized by a dominant contribution from higher harmonic responses. Further analysis of the associated vortex dynamics in the shear layer demonstrates that the intensity of these vortical structures directly governs the strength of the observed higher harmonics, and thereby the altered flame behavior.
\end{abstract}

\end{frontmatter}



\section{Introduction}
\label{sec1}

The influence of flow non-uniformities on laminar flame dynamics has long been recognized as a central problem in combustion science, owing to its dual importance in advancing fundamental understanding and improving practical models of turbulent combustion. Premixed turbulent flames are often conceptualized as ensembles of elemental laminar flame structures, or flamelets, whose collective response dictates the global flame behavior \citep{peters1988laminar}. 
Since turbulence is often interpreted as a superposition of localized flow disturbances, understanding how these disturbances affect the dynamics of individual flamelets is crucial for developing predictive and physically consistent models of turbulent combustion. Thus, the behavior of premixed flames under the unsteady and non-uniform disturbances of the surrounding flow remained a topic of interest. 

Flow non-uniformities at the upstream of a propagating premixed flame, measured and quantified as strain rates ($\textit{K}$), affect its burning rate and temperature, especially for non-equidiffusive mixtures with non-unity Lewis numbers, $Le$. 
Lewis number is defined as the ratio of the thermal diffusivity of the mixture to the mass diffusivity of the deficient species \citep{law2010combustion}.  
This recognition of flow-flame interaction mechanisms motivated a series of investigations into the behavior of non-equidiffusive ($Le\neq 1$) mixtures, particularly using counterflow and stagnation flow setups, where the dynamics of a laminar flame under steady strain rate can be studied without other effects. 
These studies paved the path for the quantitative insights into how differential diffusion influences key flame properties, including propagation speed, flammability limits, and extinction characteristics \citep{law1988propagation,kitano1986flammability}. 
It has been established that premixed flames are weakened with increasing positive strain rates if $Le > 1$, strengthened if $Le < 1$, and remain invariant to strain rate for $Le=1$ \citep{law1994structural}. 

Along with non-uniformity, practical operations also impose unsteadiness in the combustion process. These flames are often subjected to broadband, multi-frequency, and multi-scale perturbations, originating from turbulence, acoustics, or fluctuations in the incoming reactant flow. The resulting flame dynamics are inherently nonlinear and complex, shaped by a broad spectrum of time and length scales \citep{sujith2021dynamical,balachandran2005experimental}. 
In particular, temporal oscillations in flow conditions can induce oscillatory strain rates, which in turn can significantly influence the structure, stability, and extinction behavior of premixed flames. 
To gain a mechanistic understanding of how flames respond to unsteady strain rates, researchers have studied canonical configurations, such as counterflow flames, under controlled periodic excitation. 
The counterflow configuration is widely regarded as a practical setup for investigating unsteady flame behavior because it provides a well-defined, quasi-one-dimensional geometry that isolates key flame-flow interactions without the geometric complexities of practical combustors. Moreover, the counterflow flame is particularly amenable to both theoretical analysis and numerical simulation, while still being realizable in laboratory-scale experiments 
The most common approach in unsteady counterflow setups involves introducing an oscillatory velocity at the nozzles, creating a time-varying strain rate on the flame. Previous studies on unsteady strain rates have shown that the dynamics of a flame in an oscillating flow depend on the amplitude and frequency of the perturbations, as well as the average strain rate. 
The overall flame response is characterized using the Strouhal number ($S=f_e t_{flow}$) and Stokes parameter ($\mathit{St}=f_e t_{F}$), which compares the perturbation timescale with the flow ($t_{flow}$) and flame ($t_{F}$) timescales, respectively \citep{bansal2012flame,egolfopoulos1996unsteady,pearlman1995extinction,welle2003response,joulin1994response}. 
For low-frequency perturbations ($\mathit{St}<<1$), the flame response is quasi-steady as it promptly responds to the introduced perturbations. However, at high frequencies ($\mathit{St}\approx1$), the flame timescale becomes comparable to the perturbation period, resulting in a complex response. At even higher frequencies, the flame response diminishes, as it cannot adjust to the rapid fluctuations in the incoming flow. Eventually, at extremely elevated frequencies ($\mathit{St}>>1$), the flame becomes effectively invariant to these oscillations 
\citep{brown1998oscillatory,im2000effects,zhang2017effect,cuoci2013extinction,zirwes2021situ}. 

Beyond flame response characteristics, prior studies have also examined extinction phenomena in unsteady counterflow flames and contrasted them with steady configurations \citep{decroix1999study,stahl1991numerical}. Some investigations quantified the number of imposed oscillation cycles required to trigger extinction, highlighting its dependence on both the oscillation amplitude and frequency \citep{sung2000structural,kistler1996extinction}. Our earlier work \citep{potnis2021extinction} demonstrated that for $Le>1$ conditions, flames subjected to oscillatory strain rates can survive at maximum strain rate significantly above the steady-state extinction threshold, indicating a possible extension of the flammability limit under unsteady excitation. This effect, however, was not observed for $Le\leq1$ conditions.

The above brief review of the literature on unsteady counterflow flames reveals significant development over the past few decades. However, almost all prior works have focused on the centerline of the counterflow flame, where transport processes can be simplified into one-dimensional representations under steady conditions. In contrast, off-center flame dynamics can yield valuable insights by capturing inherently multidimensional effects. This off-center perspective becomes particularly important for unsteady dynamics, as often shown in turbulent counterflow setups, an extreme case of unsteadiness, where analyses naturally encompass both central and off-center regions \citep{pereira2025extinction,tirunagari2017characterization,yoo2005characteristic,jozefik2015one}. 

Motivated by these gaps, this study examines an unsteady laminar counterflow flame, with a focus on both centerline and off-center regions. Employing high-speed imaging, Mie-scattering, and Particle Image Velocimetry (PIV), we first characterize the contrasting flame dynamics at near- and off-center locations. 
Since flame response is intrinsically linked to the surrounding flow field and its hydrodynamics, this work focuses on examining the behavior of a canonical flame as the inherently three-dimensional counterflow varies and transitions toward unsteady conditions. We uncover the governing mechanisms that dictate the imposed oscillation and strain rates across different radial positions. Particular attention is given to the role of vortex dynamics and vortex shedding, which emerge due to the different excitation conditions imposed, and play a critical role in shaping flame dynamics.


\section{Methods} \label{sec:methods}
\subsection{Experimental setup}
\begin{figure}[]
     \centering
     \includegraphics[width=\columnwidth]{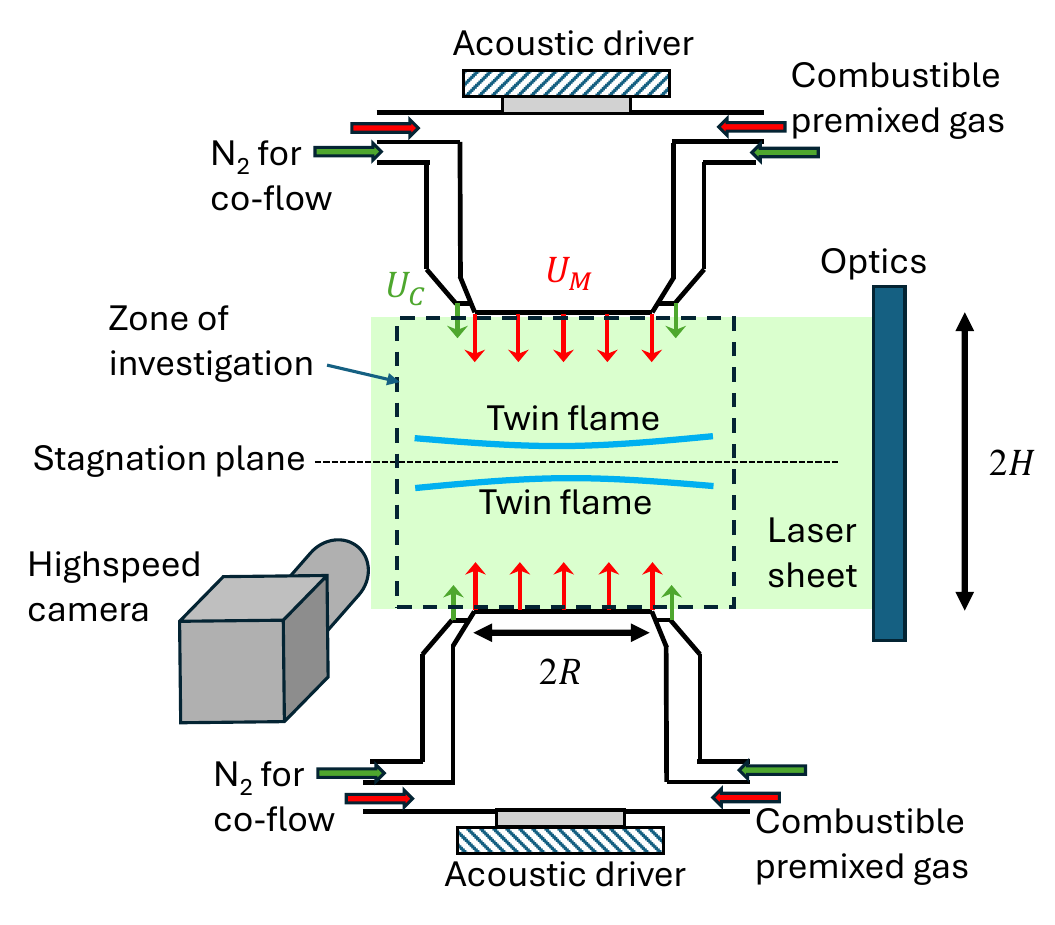}
    \caption[]{\small Illustration of the primary components of counterflow setup along with diagnostic instruments.}   
    \label{fig:Setup}  
\end{figure}

This study uses a counterflow burner with two identical $15$~mm diameter nozzles with a separation distance ($2H$) of $20$~mm. The burner is operated in the twin flame configuration (as shown in Fig. \ref{fig:Setup}), where symmetrical stable flames are established on each side of the stagnation plane by supplying premixed gas at equal flow rates through both nozzles. 
To produce perturbations in the flow through the nozzles, two speakers (acoustic drivers) are mounted at the end of the plenum chambers on each side of the counterflow setup, as shown in Fig. \ref{fig:Setup}. The speakers are driven by a voltage signal generated using the NI SignalExpress software and supplied using NI-DAQmx and an amplifier. The speakers are calibrated before the experiments to assess the level of perturbation produced based on a given input amplitude and frequency of the voltage. 
A co-flow of nitrogen ($\mathrm{N_2}$) is used to envelop the premixed gas and flame, effectively isolating them from the surrounding environment. This approach is employed to mitigate any potential impact of the ambient air on flame behavior. 

\subsection{Mixture condition}
Since the focus of this study is primarily to assess the role of flow dynamics on the flame in an unsteady counterflow setup, we maintain the conditions for the premixed flames unchanged. 
In experiments, propane ($\mathrm{C_3H_8}$), oxygen ($\mathrm{O_2}$), and nitrogen ($\mathrm{N_2}$) are supplied using three adjustable Alicat mass flow controllers (Fuel: full range of 0-10~SLPM; $\mathrm{O_2}$ \& $\mathrm{N_2}$: full range of 0-100~SLPM, both with $\pm1\%$ uncertainty) to produce a premixed gas with unity equivalence ratio ($\phi~$= 1), and an oxygen percentage ($[\mathrm{O_2}]/[\mathrm{O_2}+\mathrm{N_2}]$) of $17.8\%$. The total flow rate of 20~SLPM is used through the two nozzles (corresponding to 10~SLPM to each side of the counterflow). The fundamental flame properties, including the adiabatic flame temperature, laminar planar flame speed, and flame thickness, are calculated using the Premixed module in Chemkin-Pro software, and the GRI-3 reduced chemistry model \citep{gri3}. The effective Lewis number of the mixture, evaluated based on the method described in Addabbo \textit{et al.} \citep{addabbo2002wrinkling}, is $Le=1.46$. 
In counterflow setups, the global strain rate imposed on the flames increases (decreases) with an increase (decrease) in flow rate through nozzles. Under steady operation, premixed counterflow flames exhibit extinction beyond a critical strain rate or flow rate. We evaluate this steady-state extinction strain rate ($K_{ext,s}$) for our mixture by performing separate experiments (See Tab. \ref{tab: flame_prop}). 
For unsteady counterflow, our previous study \citep{potnis2021extinction} demonstrated that for $Le > 1$, flames can sustain instantaneous strain rates greater than the steady state extinction values.
To avoid the effects of these near-extinction dynamics, this study is conducted with flow rate or strain rate far from the steady state and unsteady extinction limits. 
The properties of a steady flame are reported in Table \ref{tab: flame_prop}. 

\begin{table*}
    \centering
    \begin{tabular}{ccccccccc}
   \hline 
 $\phi$ & ${[\mathrm{O_2}]}/{[\mathrm{O_2}+\mathrm{N_2}]}$ & $T_b$ & $S_{L,u}$ & $\delta_{L}$ & $t_F$ & Le & $K_{ext,s}$ &  $\langle{K}\rangle$\\ 
   & \% & (K) & (cm/s) & (cm) & (ms) &  & (1/s)  & (1/s) \\ 
\hline
\hline
 \\
 1.0 & 17.8 & 2000 & 31.5 & 0.045 & 1.43 & 1.46 & 640 & 180\\ 
 \hline
    \end{tabular}
    \caption{Properties of the used mixtures of propane ($\mathrm{C_3H_8}$), oxygen ($\mathrm{O_2}$), and nitrogen ($\mathrm{N_2}$). $\phi$: equivalence ratio, $T_b$: adiabatic flame temperature, $S_{L,u}$: laminar planar flame speed, $\delta_{L}$: laminar planar flame thickness, $t_F$: flame timescale defined as $\delta_L/S_{L,u}$, $Le$: effective Lewis number, calculated using the formulation by Addabbo \textit{et al.} \citep{addabbo2002wrinkling}, $K_{ext,s}$: extinction stretch rate under steady condition, and $\langle{K}\rangle$: mean strain rate used for this study. Here, global strain rate is defined as $K=2U/H$ \citep{niemann2015accuracies}.}
    \label{tab: flame_prop}
\end{table*}

\begin{figure}[h]
     \centering   
     \includegraphics[width=1\columnwidth]{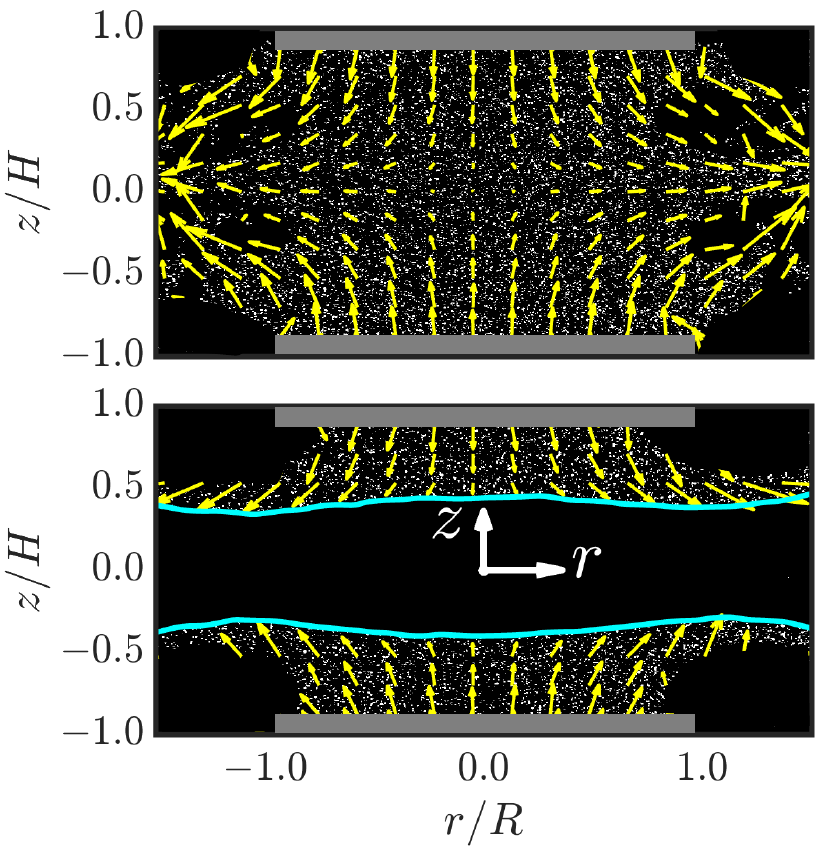}
    \caption[]%
    {{\small Mie-scattering with superimposed PIV velocity field snapshot, for non-reactive (top) and reactive (bottom) data for $f_e$ = 137~Hz. Flame edge is illustrated in cyan, nozzle exits are depicted by grey rectangular boxes, and the coordinate system is illustrated in white in the bottom panel.}}    
    \label{fig:Mie_PIV}  
\end{figure}

\subsection{Diagnostic techniques}

Mie-scattering imaging and Particle Image Velocimetry (PIV) are employed to visualize the dynamics of the flame front and measure the flow velocities. The premixed gas emanating from the nozzles is seeded with Di-Ethyl-Hexyl-Sebacate (DEHS) droplets with a diameter of around $1 - 2~\mu$m. A high-speed dual-head Nd-YLF laser and associated optics are used to produce a vertical 2D laser sheet (Sheet thickness: FWHM: $\sim 400~\mu$m). The laser sheet is placed between the nozzles and aligned diametrically as shown in Fig. \ref{fig:Setup}. 
A Phantom V710 high-speed camera, positioned orthogonally to the laser sheet and synchronized with the laser pulses, records the Mie-scattering images. The camera is connected to a $27$~mm extension tube, a Nikon Teleconverter ($2~\times$ magnification), and an AF-S Nikkor $28-70$~mm lens. The images are captured 
using a $928 \times 624$ pixel$^2$ window size, where the spatial resolution is approximately $0.03$~mm per pixel. Frame-straddled mode with a time interval of $100~{\mu}s$, is selected to facilitate appropriate particle displacements between correlated image pairs, limiting errors in PIV measurements. The data acquisition rate is maintained at $10,000$ frames per second with a total recording time of 1.78~s. 
The 2D PIV algorithm of LaVision's DaVis software is employed to obtain the velocity vectors from the correlated images. For the processing, the final interrogation window size of $24\times24$ pixel$^2$ with a $75\%$ overlap is chosen, resulting in approximately $180~\mu$m spacing between PIV vectors. 
For a comprehensive analysis of the results, we conduct PIV experiments under both flame (hot flow) and no-flame (cold flow) conditions. Since the DEHS droplets evaporate near the flame, Mie-scattering images display a clear boundary (Fig. \ref{fig:Mie_PIV}), which is tracked using custom Matlab image processing codes. The identified boundary is essentially the isothermal surface at which the droplets evaporate (close to the saturation temperature of DEHS) and is denoted the flame front or flame edge. This method has been previously used successfully to analyze the geometry and statistics of flame fronts \citep{chaudhuri2015flame, liu2021jfm, upatnieks2004cnf, weinkauff2013eif, liu2021local}. 
The concentration of the seeding droplet is small, and as such, its evaporation should not have a significant effect on the flow field \citep{potnis2021extinction}. 

\begin{table*}
    \centering
    \begin{tabular}{ | m{1.7cm} || m{1.7cm}| m{1.7cm} | m{1.7cm} | m{1.7cm} |}
 \hline
 Method & $f_e=47~$Hz & $f_e=137~$Hz & $f_e=227~$Hz & $f_e=317~$Hz \\ 
\hline
\hline
 Constant Amplitude & $\overline A=0.33$ & $\overline A=0.33$ & $\overline A=0.33$ & $\overline A=0.33$ \\ 
\hline
 Constant Power & $\overline A=0.33$ & $\overline A=0.56$ & $\overline A=0.72$ & $\overline A=0.85$ \\ 
 \hline
    \end{tabular}
    \caption{Non-dimensional amplitudes ($\overline A$) of velocity perturbation used for various frequencies ($f_e$) for two different methods, constant amplitude, and constant power.}
    \label{tab: Amp_values}
\end{table*}


\subsection{Experimental conditions}
As mentioned before, we use a fixed total flow rate ($20$ SLPM) for this study, which results in an averaged flow velocity through each nozzle of $U = 0.9$~m/s, corresponding to a Reynolds number of $Re = 2\rho RU/\mu = 900$, based on nozzle radius ($R=7.5\times10^{-3}~$m), weighted density value ($\rho=1.2~$kg/m$^3$), and weighted dynamic viscosity value of the premixed gas ($\mu = 1.8 \times 10^{-5}~$N$\cdot~$s/m$^2$) at inlet condition of $\sim 25^o$ C. 
Using the acoustic driver, we impose flow oscillations with various frequencies and amplitudes. The inlet flow velocity under the perturbation can be expressed as 
\begin{equation}
   U_M=U(1 + \overline A\sin{(2\pi f_et)}),
   \label{Eq:U_inlet}
\end{equation}
$f_e$ is the excitation frequency of the velocity oscillation, and $\overline A$ is the non-dimensional amplitude of the perturbation. 
For this study, single-frequency perturbations are used. The range of frequencies is chosen to ensure that the flame exhibits a measurable response to the imposed oscillation, and the acoustic driver can produce a perturbation of sufficiently strong amplitude. 
At very high frequencies or high Stokes number conditions ($\mathit{St}$), premixed flames do not exhibit a strong response \citep{brown1998oscillatory,zhang2017effect}. On the other hand, the used acoustic drivers and amplifiers exhibit irregularities and high uncertainty for frequencies smaller than 40 Hz. With these constraints, the four excitation frequencies selected for the experiments are 47, 137, 227, and 317 Hz. 

Two strategies are employed during experiments conducted across various frequencies. 
First, we employ a \textit{constant amplitude}, $\overline A=0.33$ for all frequencies, enabling us to explore the role of changing frequency on the ensuing dynamics. While \textit{constant amplitude} is a popular method, we acknowledge that it does not maintain a constant power of the perturbed flow across various frequencies. From Eq. \ref{Eq:U_inlet}, it can be shown that the rate of energy injected into the flow per unit cycle of oscillation is $(1/2)(\pi \rho R^2U)(\overline AU)^2$. Thus, the total amount of energy added per unit time or power can be defined as $\dot{E}(f_e) \approx (1/2)\pi \rho R^2\overline A^2U^3/f_e$. Since, in our experiments, $U$ and $R$ are fixed, the \textit{constant amplitude} (fixed $\overline A$) approach will have progressively lesser power at higher frequencies. Thus, to complement the \textit{constant amplitude} approach, we also explore the \textit{constant power} method, where the amplitudes for various frequencies were changed as $\overline A(f)\propto \sqrt{f_e}$, ensuring power ($\dot{E}(f_e)$) remains constant. Table \ref{tab: Amp_values} lists the amplitude values used at different frequencies for these two methods. 

\subsection{Data processing}
Before we present the results, we briefly discuss the variables of interest and the conventions used to refer to them. 
The Mie-scattering images are post-processed using edge detection techniques to extract the instantaneous flame edges. We define a flame location function, $z_f(r,t)$, which is the vertical distance between the flame edge and the stagnation plane. From 2D PIV post processing, we obtain radial, $V_r(r,z,t)$, and streamwise or axial, $V_z(r,z,t)$, components of velocities. In our analyses, we decompose time-varying variables (e.g. $\alpha(r,t)$) into their mean, denoted by $\langle \alpha\rangle(r)=(1/n)\sum_n{\alpha(r,t)}$ and fluctuation, denoted by $\alpha'(r,t)$, such that $\alpha(r,t)=\langle{\alpha}\rangle(r)+\alpha'(r,t)$. Here, $n$ is the length of the time series data. The degree of fluctuation is measured by its root mean square (RMS), defined as $\alpha|_{RMS}=\sqrt{(1/n)\sum_n(\alpha')^2}$. 
For frequency-based analyses, we use power spectral density (PSD), where $\hat{P}_\alpha$ denotes the PSD of variable $\alpha$.

\begin{figure}[t]
     \centering
     \includegraphics[width=01\columnwidth]{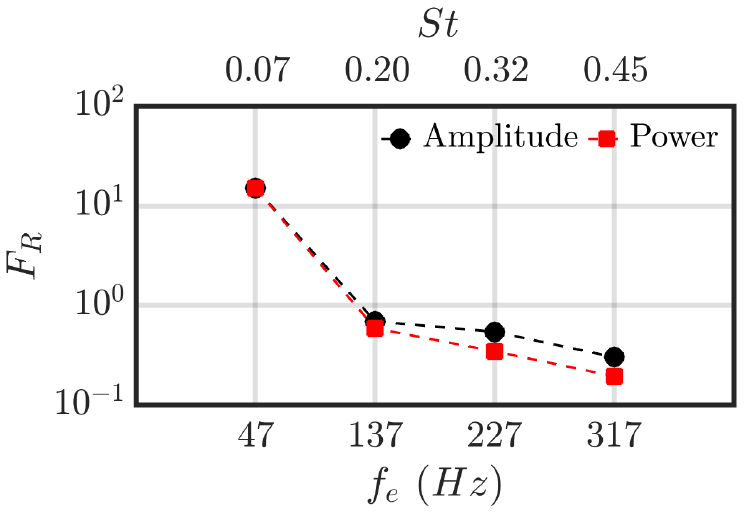}
    \caption[]%
    {\small Flame response along the center line ($r/R=0$) as a function of excitation frequency for both perturbation strategies: the \textit{constant amplitude} method in black and the \textit{constant power} method in red. The corresponding values of the Stokes number, $\mathit{St}=f_e t_F$ are shown on the secondary x-axis.}   
    \label{fig:F_R}  
\end{figure}

\subsection{Uncertainties}
The uncertainty in the flame location, $z_f$, arises from the edge detection technique. This uncertainty was evaluated by assessing the sensitivity of the identified flame location when the parameters in the edge-detection procedure in Matlab were varied by $\pm 50\%$. The statistical analyses showed a maximum of $\pm 2$ pixel uncertainty in $z_f$. 
The uncertainty in the radial and streamwise velocity components, $V_r(r,z,t)$ and $V_z(r,z,t)$, was estimated directly from DaVis software. This uncertainty arises from seeding density, correlation peak, and the chosen PIV window sizes. With the PIV setting used in this study, the expected uncertainty in velocity measurements is $ 0.01-0.02$ m/s across the measurement domain. 
The uncertainty in the stretch rates depends on the combined uncertainty of flame location, PIV velocity fields, and the error associated with velocity gradients. The estimated uncertainty for stretch rates is in the range $0.01-0.02~$1/s.




\section{Results}

\subsection{Flame response at center}
To begin our analysis, we focus on flame response measured by oscillation in flame location at various frequencies. Since the observed flame oscillation is multi-modal (shown later), we evaluate the total energy in the flame displacement along the centerline,  $E_{z_f(r/R=0)}=\int_{0^+}^{\infty}\hat{P}_{z_f(r/R=0)}df$, where, $\hat{P}_{z_f}$ is the power spectral density (PSD) of $z_f(r,t)$, the time series of vertical distance of flame edge from the stagnation plane at normalized radius $r/R$. 
Subsequently, the flame response is defined as the ratio of energies of flame displacement and imposed velocity perturbation measured at the centerline, $F_R=((E_{z_f}/ t_F^2)/{V_z|_{RMS}}^2)_{r/R=0}$, where $t_F$ is the flame time scale. The comparison shows that the flame response ($F_R$) becomes weaker with increasing excitation frequency ($f_e$) or Stokes number ($\mathit{St}$), as shown in Fig. \ref{fig:F_R}. At higher $\mathit{St}$, the perturbation timescale ($\sim1/f_e$) decreases while the flame response timescale ($\sim\delta_L/S_{L,u}$) remains fixed, thereby allowing the flame relatively less time for response. This behavior has also been reported in the literature for various flame configurations \citep {egolfopoulos1996unsteady, kistler1996extinction, borghesi2009dynamic}.  

\begin{figure*}[h]
     \centering
    \includegraphics[width=\textwidth]{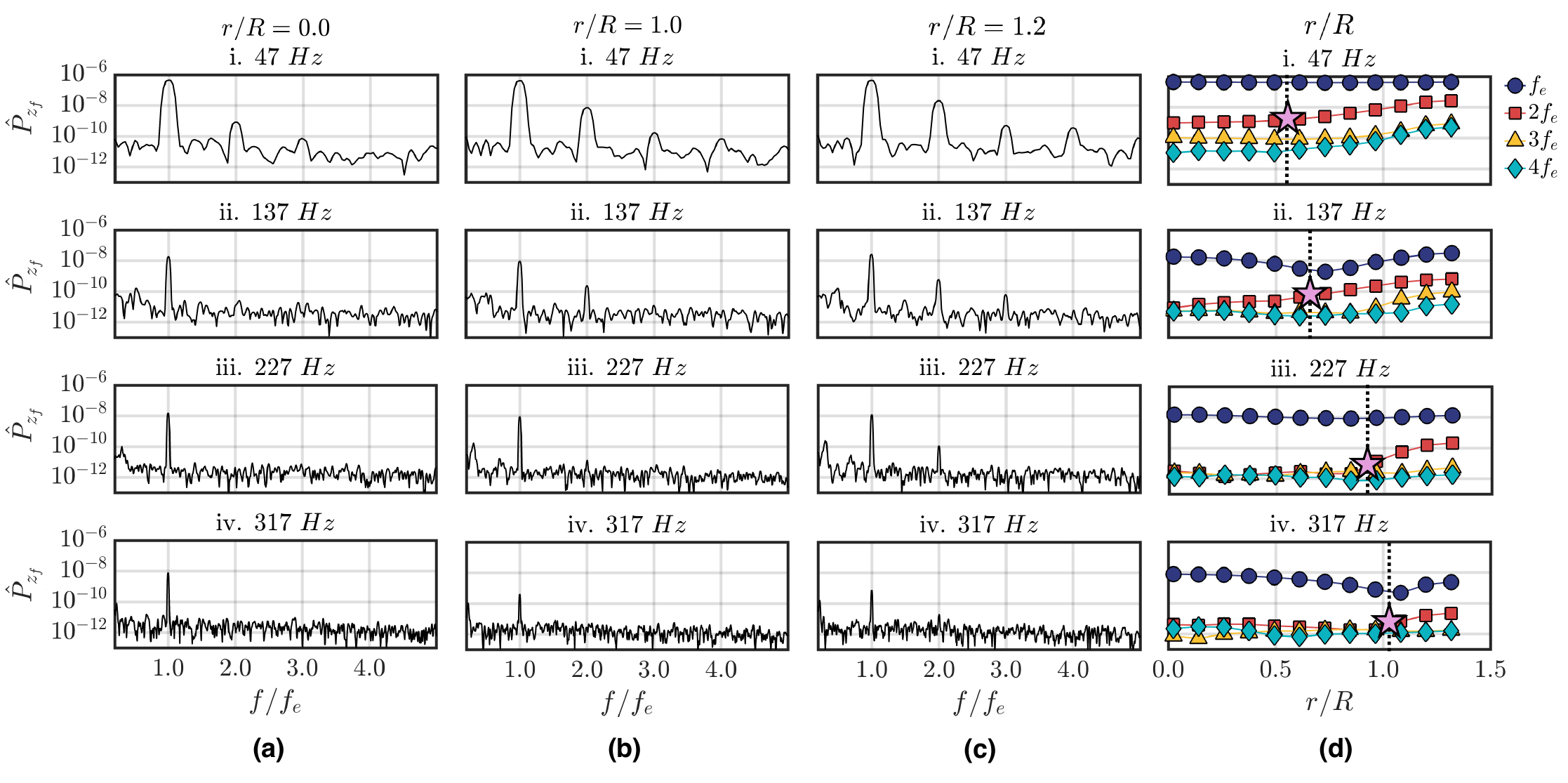}
    \caption[]%
    {{\small Power spectral density of flame oscillation or displacement ($\hat P_{z_f}$) measured at (a) flame's center point ($r/R=0$), (b) flame location $r/R=1.0$, and (c) flame location $r/R=1.2$ accompanied by (d) radial profiles of first four harmonics of the $\hat P_{z_f}$ (markers are displayed every 6 data points for clarity). The radial location ($r_{2f_e}$) where $2f_e$ starts to increase is marked with a star and a vertical dashed line in (d). Four rows represent four excitation frequencies: (i.) $47~$Hz, (ii.) $137~$Hz, (iii.) $227~$Hz, and (iv.) $317~$Hz. The uncertainty in the computation of $z_f$ is $\pm 2$ pixel.}}    
    \label{fig:FFT_zf_harm}  
\end{figure*}

Figure \ref{fig:F_R} further demonstrates that the normalized flame response exhibits similar behavior under both perturbation strategies, the \textit{constant amplitude} method and the \textit{constant power} method. This consistency highlights the importance of adopting a proper definition and normalization when comparing different conditions. 

Based on this similarity between the two strategies, to increase clarity and consistency, and to avoid repetition, we limit our discussion in the remainder of the manuscript to the results from the \textit{constant amplitude} method. The corresponding results for the \textit{constant power} approach, which corroborates the conclusions drawn from the \textit{constant amplitude} method, are provided in the Supplementary Material. 
We will also focus our analyses only on the top flame in our twin flame configuration. 
Supplementary Material includes key results from the bottom flame, which display similar flame dynamics comparable to those observed in the top flame.

\subsection{Flame response at off-center locations}

In this section, we discuss the local spectral signature of flame oscillation at both center and off-center locations. Figure \ref{fig:FFT_zf_harm}(a-c) presents the PSDs of the flame displacement ($\hat{P}_{z_f}$) at three radial locations: the centerline ($r/R=0$), and two off-center positions ($r/R=1.0$, and $r/R=1.2$), for all four excitation frequencies. The corresponding time series are reported in the Supplementary Material. 
At the centerline, the PSD displays a strong peak at the excitation frequency ($f/f_e=1$). As we move away from the center, the flame response becomes increasingly non-sinusoidal (see Supplementary Material), accompanied by the emergence of higher harmonics at integer multiples of the excitation frequency ($f/f_e=2,3,4$). 
The gradual emergence of higher frequency response is further highlighted in Fig. \ref{fig:FFT_zf_harm}(d), which compares the radial distribution of the spectral energy at the excitation frequency (dark blue circle) and its higher harmonics (red square, yellow triangle, and cyan diamond) for the different excitation cases. Across all conditions, the response near the centerline (approximately $r/R < 0.5$) is dominated by the excitation frequency, marked by high energies at $f_e$, with minimal energy content in higher harmonics ($2f_e$, $3f_e$, $4f_e$). 
As we move to outer radii (higher $r/R$ values), we observe a gradual increase in the higher harmonics, which are strongest at the extreme location of our measurement ($r/R\approx1.4$). Additionally, from the radial profiles of $2f_e$, we can identify a distinct location (noted as $r_{2f_e}$) where a gradual increase in energy is observed. These points are represented by stars in Fig. \ref{fig:FFT_zf_harm}(d)  for each excitation frequency.

\begin{figure*}
     \centering
    \includegraphics[width=\textwidth]{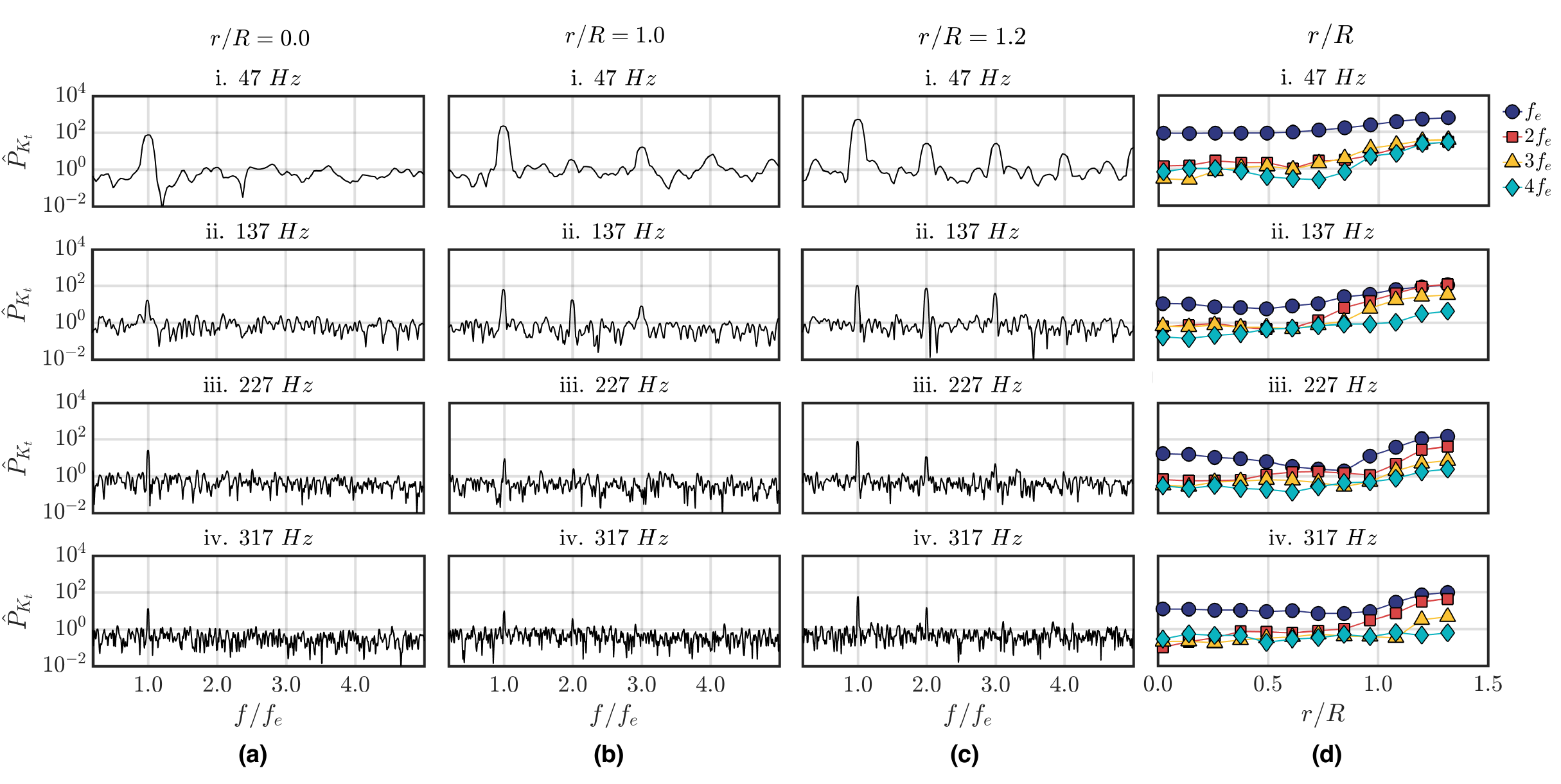}
    \caption[]%
    {{\small Power spectral density of flame conditioned stretch rate due to tangential strain ($\hat P_{K_t}$) measured at (a) flame's center point ($r/R=0$), (b) flame location $r/R=1.0$, and (c) flame location $r/R=1.2$ accompanied by (d) radial profiles of first four harmonics of the $\hat P_{K_t}$ (markers are displayed every 6 data points for clarity). Four rows represent four excitation frequencies: (i.) $47~$Hz, (ii.) $137~$Hz, (iii.) $227~$Hz, and (iv.) $317~$Hz. The uncertainty in the computation of $\overline {K_t}$ is $0.01-0.02~$1/s.}}    
    \label{fig:psd_Kt_harm}  
\end{figure*}

\subsection{Dynamics of stretches}
Next, we focus on the stretch rates the flame experiences at different radial locations. The stretch rate, defined as the fractional rate of change of flame surface area ($\textbf{K}=(1/A)dA/dt$), can be expressed as the combination of stretch rates due to pure curvature $K_c$, normal strain rate $K_n$, and tangential strain rate $K_t$. It can be shown that both $K_c$ and $K_n$ are strongly dependent on the flame curvature \citep{matalon2009flame}. For steady counterflow flames, the curvature effects ($\kappa$) are generally negligible owing to a nearly flat flame shape. On the other hand, the tangential strain is dominant due to the strong radial gradient of the flow along the flame \citep{law2010combustion}. Thus, we focus our discussion only on the stretch due to tangential strain, $K_t=\nabla_t\cdot\textbf{v}_t={\partial V_t}/{\partial \eta}$, which represents the tangential gradient of the tangential component of the flow velocity at the flame edge. Here, $\eta$ is the tangential direction along the flame.

Figure \ref{fig:psd_Kt_harm}(a-c) shows the power spectral density of flame conditioned stretch rate due to tangential strain, $\hat{P}_{K_t}$, at three radial locations: the centerline ($r/R=0$), and two off-center positions ($r/R=1.0$, and $r/R=1.2$), for all four excitation frequencies. Figure \ref{fig:psd_Kt_harm}(d) compares the radial distribution of the spectral energy at the excitation frequency and its higher harmonics for the different excitation cases. Clearly, PSD of $K_t$ bears a strong resemblance to the PSD of flame oscillations. Similar to $\hat{P}_{z_f}$ (shown in Fig. \ref{fig:FFT_zf_harm}), $\hat{P}_{K_t}$ shows a strong signature of the excitation frequency near the center of the flow, while higher harmonics become more energetic at the outer edges of the flow. Although not discussed (shown in Supplementary Material) here, PSDs of $K_c$, and $K_n$ also demonstrate similar behavior. 

\begin{figure}[t!]
     \centering
    \includegraphics[width=\columnwidth]{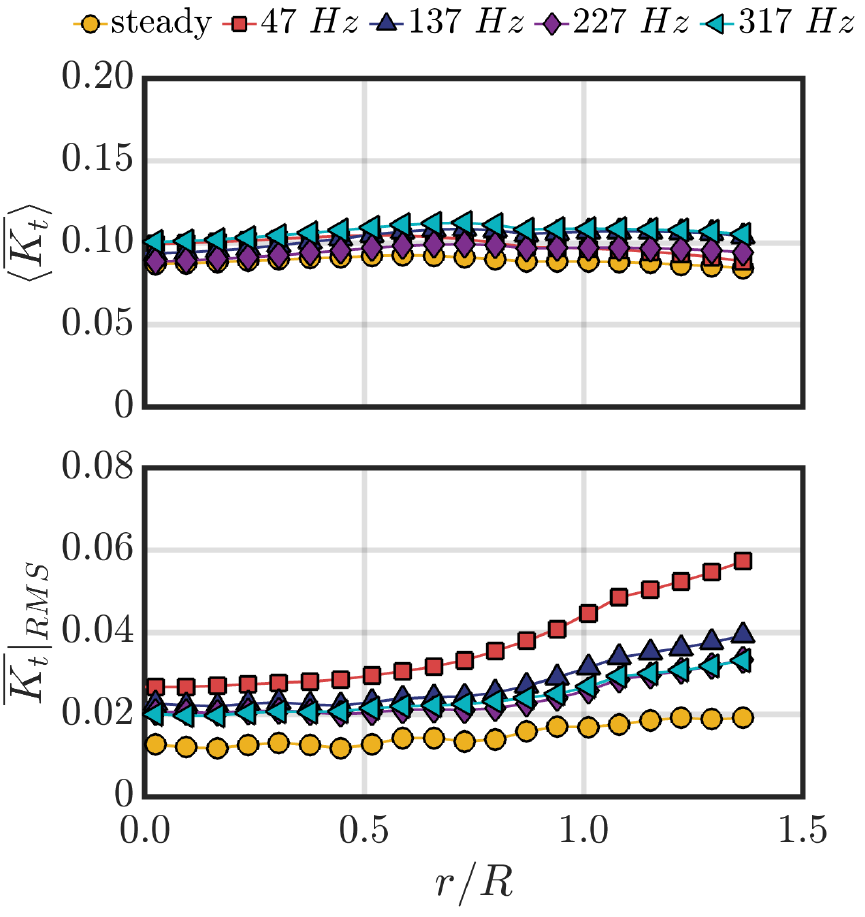}
    \caption[]%
    {{\small Radial profiles of mean and root mean square of normalized stretch rate due to tangential strain, $\overline {K_t}$, at different excitation frequencies and for the steady case (markers are displayed every 3 data points for clarity).}} 
    \label{fig:Kt_mean_RMS}  
\end{figure}

We have also extracted the radial profiles of the mean, $\langle \overline{K_t} \rangle$, and RMS of the fluctuations, $\overline{K_t}|_{RMS}$ for the normalized stretch rate due to tangential strain ($\overline{K_t}=K_t t_f$) under four excitation frequencies and compare them with the steady condition in Fig. \ref{fig:Kt_mean_RMS}. 
The mean tangential strain does not exhibit significant variations along the radius, a hallmark of a counterflow or opposed-jet configuration. Furthermore, since the mean flow rate or axial velocity ($U$) is kept constant across all steady and unsteady conditions, the mean tangential strain remains nearly constant across all excitation conditions. 
The RMS, on the other hand, increases significantly from the steady to unsteady condition (e.g., $f_e=47$~Hz) due to the imposed oscillation in $U_{M}$. We also notice a heightened $\overline{K_t}|_{RMS}$ with radius, originating from the increased degree of oscillation of higher harmonics at the outer radii, as evident in Fig. \ref{fig:FFT_zf_harm}(d). The $\overline{K_t}|_{RMS}$, however, decreases with increasing excitation frequency, due to weaker flame oscillations, evident in Fig. \ref{fig:F_R}.

\subsection{Discussion}
The above results show some interesting behavior in unsteady counterflow, particularly at flame locations away from the center. Near the center of the flame, the flame response (both $z_f$ and $K_t$) is primarily dominated by the excitation frequency. This was also reported in previous studies on unsteady counterflow flames, which focused their interest only on the centerline dynamics \citep{egolfopoulos1996unsteady,welle2003response}. Our results, however, unveil a different behavior at the outer edges of the flame. In particular, we notice a gradual strengthening of the second harmonics, as well as third and fourth harmonics, as we move radially outward. 
These observations naturally raise some new questions that were not addressed in previous studies. They are:

\begin{enumerate}[label=Q\arabic*:]
    \item Why do the higher harmonic oscillations emerge at outer radii of unsteady counterflow flames?

    \item Is there a critical radius where we expect the higher harmonics to become important?

    \item Why do certain excitation frequencies induce multiple higher harmonics in flame oscillations at the outer flame radii (e.g, $f_e =47~$Hz shows 2nd, 3rd, and 4th harmonics), while others result in fewer high harmonics (e.g, $f_e =227~$Hz shows only 2nd harmonic)?
    
\end{enumerate}

To address these questions and provide a broader perspective on unsteady counterflow phenomena, next, we analyze the flow field measured using PIV. We recognize that the flame responds to the underlying flow field, hence, we present and analyze the PIV data from cold-flow (no flame) experiments. To ensure that the flow and dynamics remain essentially unchanged in the presence of the flame, some PIV data from the hot-flow (with flame) experiments are included in the Supplementary Material. The cold-flow and hot-flow PIV data show both qualitative and quantitative similarities, albeit with some limitations in the latter due to the lack of seeding in some regions of the flow. 

\begin{figure*}
     \centering
     \includegraphics[width=0.8\textwidth]{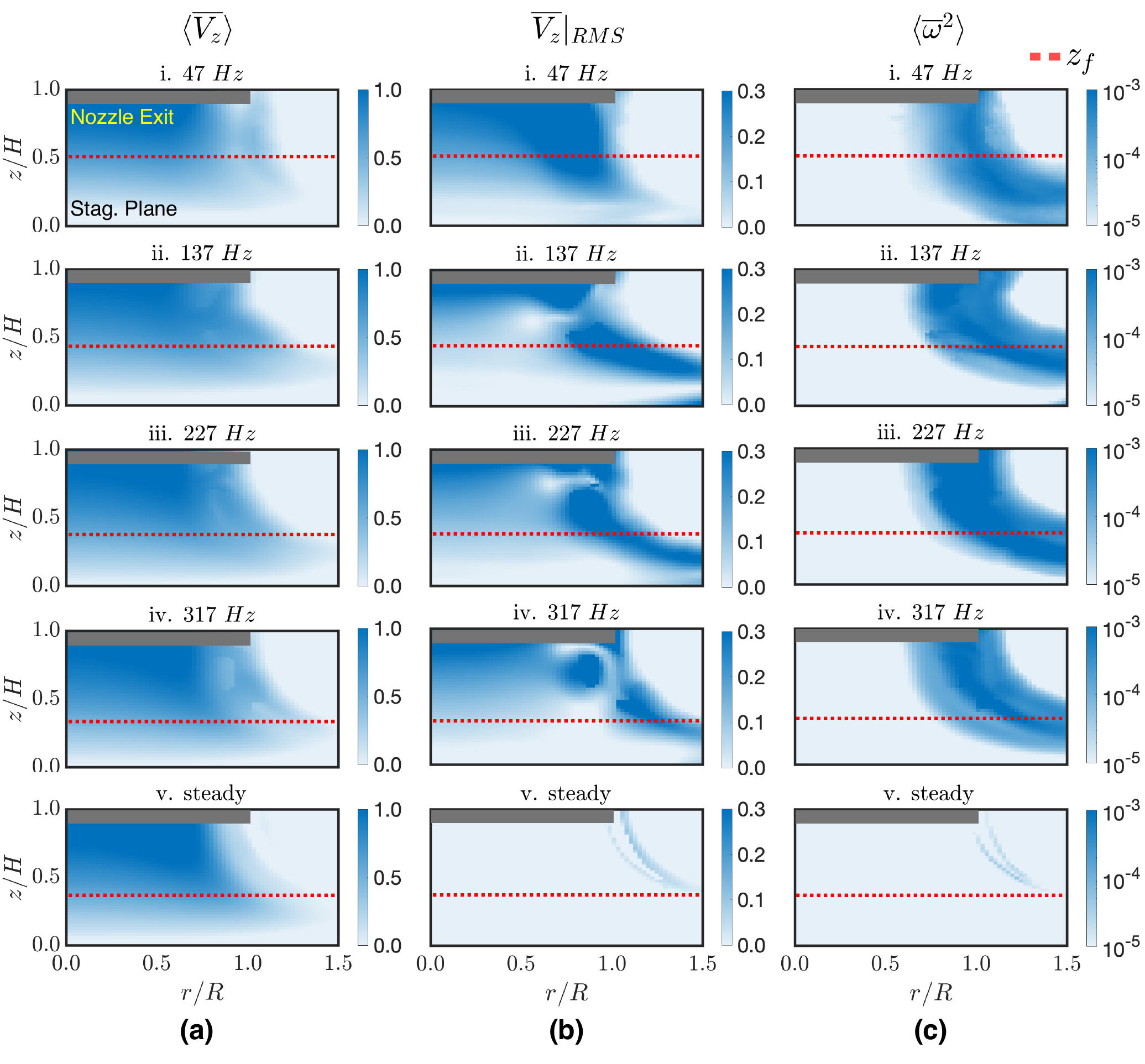}
    \caption[]%
    {{\small Color maps for the distribution of (a) normalized stream wise mean velocity, $\langle \overline{ V_z} \rangle$, (b) normalized stream wise root mean square velocity, $\overline{V_z}|_{RMS}$, and (c) normalized mean of vorticity squared, $\langle \overline \omega^2 \rangle$ at different excitation frequencies (i.) $47~$Hz, (ii.) $137~$Hz, (iii.) $227~$Hz, and (iv.) $317~$Hz, and for the steady (non-oscillated) case (v.). The mean flame location $\langle z_f\rangle$ for each frequency case is depicted with a red dashed line. Color maps represent the field to the right of the counterflow vertical centerline and above the stagnation plane. The uncertainty in the computation of $V_z$ is $0.01-0.02~$m/s.}}    
    \label{fig:Vz_Vzrms_w_wrms_fe}  
\end{figure*}

\subsubsection{Velocity and vorticity fields}
We begin by examining the normalized axial (streamwise) velocity ($\overline{V_z}=V_z/U$) distribution, focusing on its mean, $\langle \overline{V_z} \rangle$, and RMS of the fluctuations, $\overline{V_z}|_{RMS}$, presented in columns (a) and (b) of Fig. \ref{fig:Vz_Vzrms_w_wrms_fe}, respectively. The results are shown for unsteady conditions at four excitation frequencies (rows i.–iv.) and for the steady (no oscillation) condition (row v.). 
Across all five cases, the mean axial velocity fields (Fig. \ref{fig:Vz_Vzrms_w_wrms_fe}a) unsurprisingly exhibit a similar pattern, as a constant mean flow rate is maintained. At the nozzle exit ($z/H=1.0$), $\langle \overline{V_z} \rangle$ is uniform near the centerline ($r/R = 0$) or core of the flow, but decays toward the outer edge ($r/R=1.0$), forming a shear layer identified by the color (dark to light) gradient in radial direction in Fig. \ref{fig:Vz_Vzrms_w_wrms_fe}(a). Furthermore, as expected for a counterflow setup, the mean axial velocity, $\langle \overline{V_z} \rangle$, decreases as we move towards the stagnation plane ($z/H=0$), where it approaches 0 for all conditions.

The RMS of fluctuations for unsteady conditions displays drastically different behavior from that of the steady operations. 
For unsteady conditions, $\overline{V_z}|_{RMS}$ is higher due to the imposed oscillations. At the nozzle exit ($z/H=1.0$), and near the centerline ($r/R=0$), $\overline{V_z}|_{RMS}$ is almost uniform across the radius for all excitation cases (rows i.–iv., Fig. \ref{fig:Vz_Vzrms_w_wrms_fe}b). Interestingly, at the outer edge of the flow ($r/R\approx 1$), along the shear layer, elevated $\overline{V_z}|_{RMS}$ is observed. 
The combination of a steady co-flow and an oscillatory main flow generates inherently unsteady velocity gradients, which amplify $\overline{V_z}|_{RMS}$ within the shear layer. 
As we move axially towards the stagnation plane ($z/H=0$), we observe that the shear layer, marked by the high $\overline{V_z}|_{RMS}$ zone, moves radially outward due to radial expansion of the flow, a hallmark of counterflow setups. Near the centerline, however, the $\overline{V_z}|_{RMS}$ diminishes at the stagnation plane. 
Quantitatively, $\overline{V_z}|_{RMS}$ has similar magnitudes across all excitation frequencies for the \textit{constant amplitude} excitation method. 
On the other hand, for the steady (no-oscillation) case, due to lack of imposed perturbation, $\overline{V_z}|_{RMS}\approx 0$, near the center or core of the flow as shown in Fig. \ref{fig:Vz_Vzrms_w_wrms_fe}(b), row v. At the shear layer, the $\overline{V_z}|_{RMS}$ is non-zero, but much weaker compared to the unsteady conditions. 

\begin{figure*}[]
     \centering
    \includegraphics[width=0.8\textwidth]{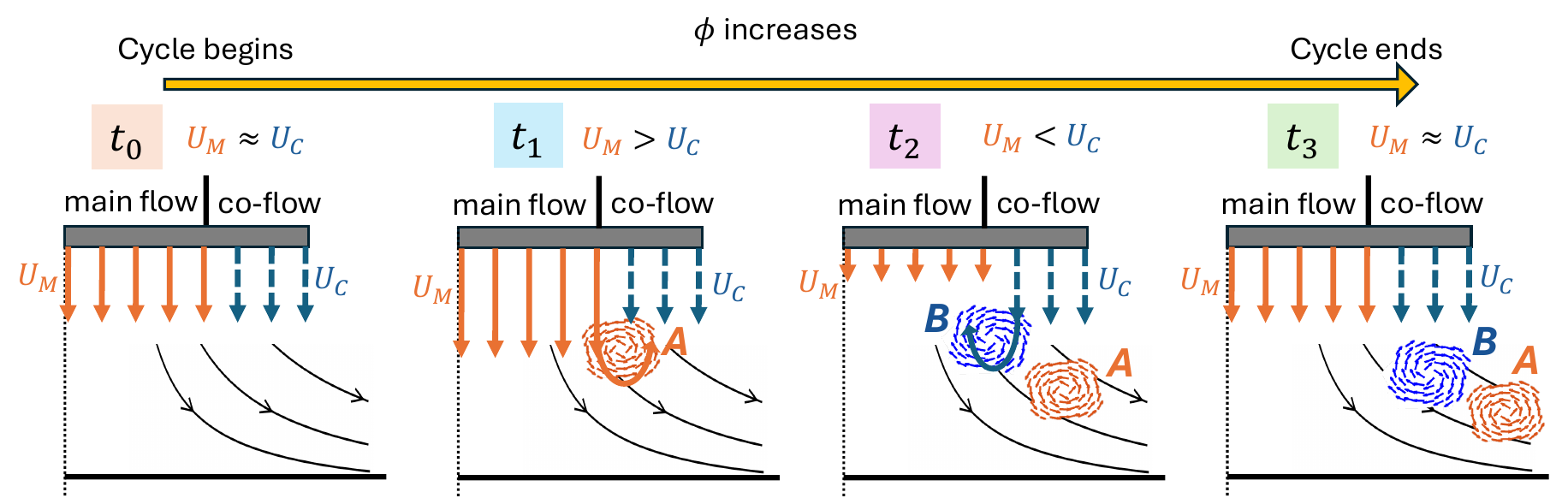}
    \caption[]%
    {{\small Schematic illustration of vortex dynamics along shear layer at different phase instances. The schematic represents the top-right quadrant corner of the counterflow.}}  
    \label{fig:Sketch_phaseAvg_w_Vz}  
\end{figure*}

\begin{figure*}[]
    \centering
    \includegraphics[width=\textwidth]{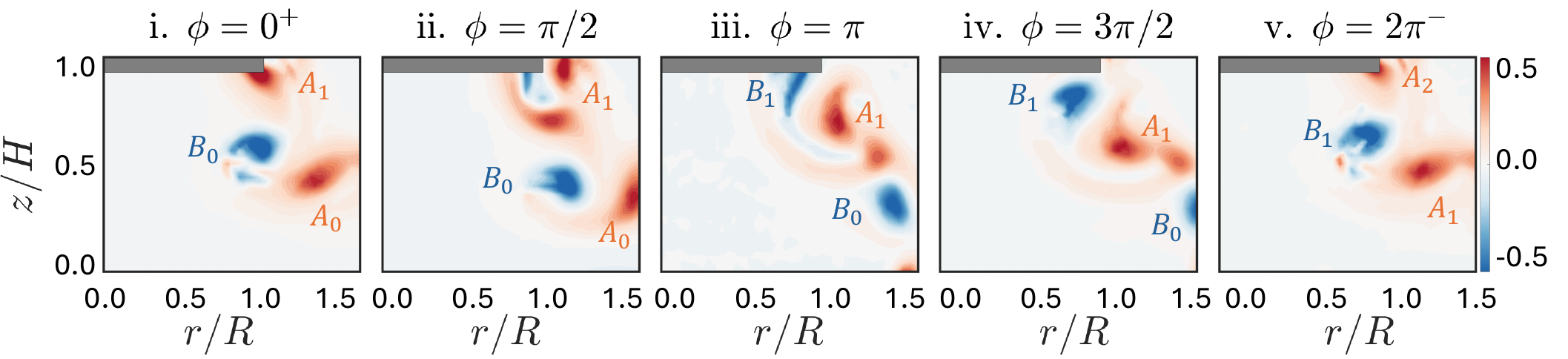}
    \caption[]%
    {{\small Phase-averaged vorticity field for $f_e=137$~Hz at five phase angles from $0^+$ to $2\pi^-$ (i.-v.). The schematic represents the top-right quadrant corner of the counterflow. A and B are large counter-rotating vortex pairs generated during each cycle. Subscripts $0$, $1$, and $2$ refer to vortices originating from the previous, current, and next cycle, respectively. }}  
    \label{fig:phaseAvg_w_Vz}  
\end{figure*}

\begin{figure*}[]
     \centering
    \includegraphics[width=0.75\textwidth]{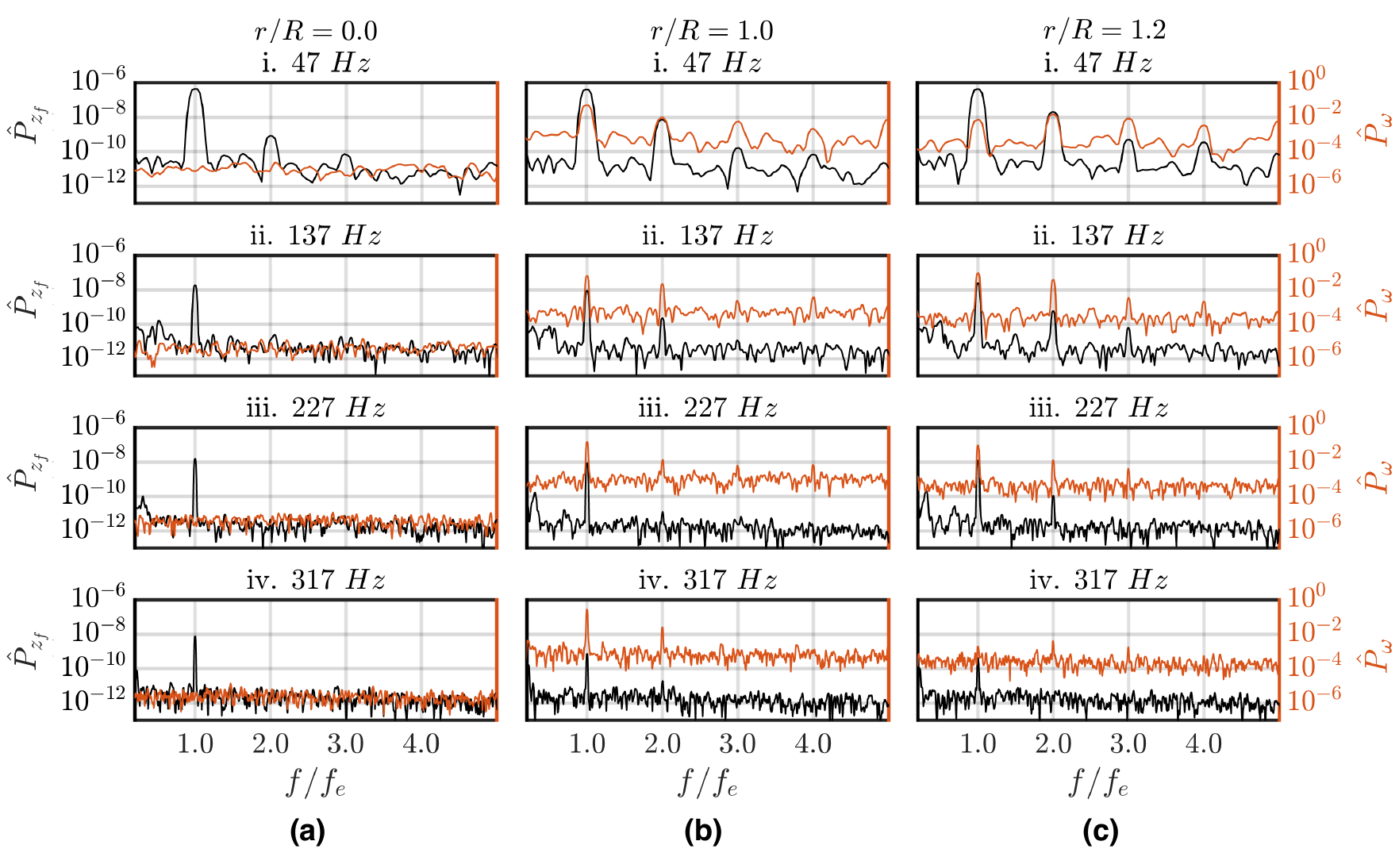}
    \caption[]%
    {{\small Power spectral density of flame displacement ($\hat P_{z_f}$) and flow vorticity ($\hat P_{\omega}$) measured at (a) center line ($r/R=0$), (b) flame location $r/R=1.0$, and (c) flame location $r/R=1.2$ at different excitation frequencies ((i.) $47~$Hz, (ii.) $137~$Hz, (iii.) $227~$Hz, and (iv.) $317~$Hz) at mean flame location.}}  
    \label{fig:psd_zf_w}  
\end{figure*}

From the preceding discussion and the $\langle \overline{V_z} \rangle$ distribution, it is clear that a well-defined shear layer develops along the outer edge of the flow. Under forced oscillation conditions, the elevated $\overline{V_z}|_{RMS}$ values within this region signify pronounced unsteadiness. This behavior is further corroborated by the normalized vorticity field, $\overline{\omega}=\omega/(U/R)$. Since $\overline{\omega}$ can assume both positive and negative values, and our primary interest lies in the strength of the vorticity field, we examine the mean of the square of the normalized vorticity, i.e., $\langle \overline \omega^2  \rangle$, in columns (c) of Fig. \ref{fig:Vz_Vzrms_w_wrms_fe}. 
Several features emerge from the vorticity distributions. The inner region (near the center and far from the shear layer) remains largely \textit{irrotational}, characterized by negligible $\langle \overline \omega^2  \rangle$. In contrast, the shear-layer region displays strong \textit{rotational} behavior, with elevated $\langle \overline \omega^2  \rangle$, signifying sustained vortex shedding. For the steady case, the $\langle \overline \omega^2  \rangle$ is much weaker, even in the shear layer, reflecting the absence of coherent periodic excitation and vortex shedding.

\subsubsection{Dynamics of shear layer}
To explore the origin of enhanced vortical activity in the shear layer of unsteady counterflow, we first consider a mechanistic description of the flow dynamics. 
We can hypothesize that the imposed oscillation causes a time-varying mismatch in velocity between the main flow and the co-flow, as illustrated in Fig. \ref{fig:Sketch_phaseAvg_w_Vz}. As explained in Sec. \ref{sec:methods}, the co-flow velocity ($U_C$) is constant and roughly equal to the mean velocity of the main flow ($U_C \approx U$), while the main flow velocity ($U_M$) follows a sinusoidal pattern following Eq. \ref{Eq:U_inlet}. At the start of the cycle (time = $t_0$, phase angle, $\phi=0$), the velocity of the main flow from the nozzle ($U_M$) approximately matches that of the co-flow ($U_C$). During the first half of the cycle ($0 < \phi <\pi$), when the main flow is faster than the co-flow, i.e. $U_M>U_C$, a counterclockwise-rotating vortex (vortex A at time = $t_1$ in Fig. \ref{fig:Sketch_phaseAvg_w_Vz}) forms at the shear layer, which becomes fully developed and convected downstream by $\phi = \pi$. 
In the second half of the cycle ($\pi < \phi < 2\pi$), the main flow becomes slower than the co-flow, i.e. $U_M<U_C$, generating a clockwise-rotating vortex (vortex B at time = $t_2$ in Fig. \ref{fig:Sketch_phaseAvg_w_Vz}). This vortex grows in strength and is advected downstream by the time the cycle reaches $\phi=2\pi$. As shown at time $t_3$, in Fig. \ref{fig:Sketch_phaseAvg_w_Vz}, these vortices will approach and interact with the outer edges of the counterflow flame.

This hypothesized description is then supported by the phase-averaged vorticity field in Fig. \ref{fig:phaseAvg_w_Vz} (for $ f_e=137$ Hz; other conditions are shown in the Supplementary Material). These phase-averaged snapshots spanning from $\phi=0^+$ to $\phi=2\pi^-$ reveal the real-time evolution and advection of the vortices along the shear layer. The sequence clearly shows the formation and advection of the pair of counter-rotating vortices (A$_1$ and B$_1$ in the figure) along the shear layer during the cycle, and their eventual approach towards the outer edge of the flame. 

Since each cycle of imposed oscillation in the main flow, two counter-rotating vortices are formed, and each of which perturbs the flame, the effective frequency of flame oscillation is expected to be twice the excitation frequency. Thus, near the shear layer ($r/R\approx 1$), the second harmonic ($f/f_e=2$) becomes the dominant component in the flame oscillation (Fig. \ref{fig:FFT_zf_harm}).


To further elucidate the vortex dynamics, we compute the power spectral density (PSD) of vorticity, $\hat{P}_\omega$, in the vicinity of the mean flame location (red dashed line in Fig. \ref{fig:Vz_Vzrms_w_wrms_fe}). Figure \ref{fig:psd_zf_w} compares $\hat{P}_\omega$ with the PSD of flame oscillation, $\hat{P}_{z_f}$, at three radial positions. Along the centerline ($r/R=0$), which lies outside the shear layer, the vorticity spectrum remains essentially flat across all frequencies (Fig. \ref{fig:psd_zf_w}a), indicating negligible vortical activity and confirming that the counterflow core is irrotational. However, near the shear layer, $\hat{P}_\omega$ exhibits distinct peaks at both the excitation frequency ($f/f_e=1$) and its second harmonic ($f/f_e=2$), as shown in Fig. \ref{fig:psd_zf_w}(b) and (c) for $r/R=1.0$ and $1.2$, respectively. Notably, the vorticity spectrum near the shear layer closely mirrors that of flame oscillation, whereas near the centerline, the two spectra differ markedly.

Now that we have identified the morphology and dynamic nature of the flow field, we return to the three questions (Q1-Q3) posed at the beginning of this section, concerning the disparate oscillation dynamics observed in unsteady counterflow.

\noindent\textbf{Q1: Why do the higher harmonic oscillations emerge at outer radii of unsteady counterflow flames? }
Based on the preceding analyses of velocity and vorticity, the dynamics of an unsteady counterflow flame can be delineated into two distinct zones. In the near-core region, the flow remains largely irrotational, and flame oscillations are governed exclusively by the externally imposed oscillation of the main flow. Consequently, $\hat{P}_{z_f}$ and $\hat{P}_{K_t}$ display a dominant peak at the excitation frequency, $f/f_e=1$. In contrast, at the outer edge of the flow, the shear layer is strongly rotational owing to the generation of two counter-rotating vortices during each oscillation cycle. Here, the flame is influenced not only by the imposed oscillation but also by the vortex dynamics, leading $\hat{P}_{z_f}$ and $\hat{P}_{K_t}$ to exhibit dominant peaks at both $f/f_e=1$ and $f/f_e=2$. 

\vspace{1pt}

\begin{figure}[t]
     \centering
    \includegraphics[width=\columnwidth]{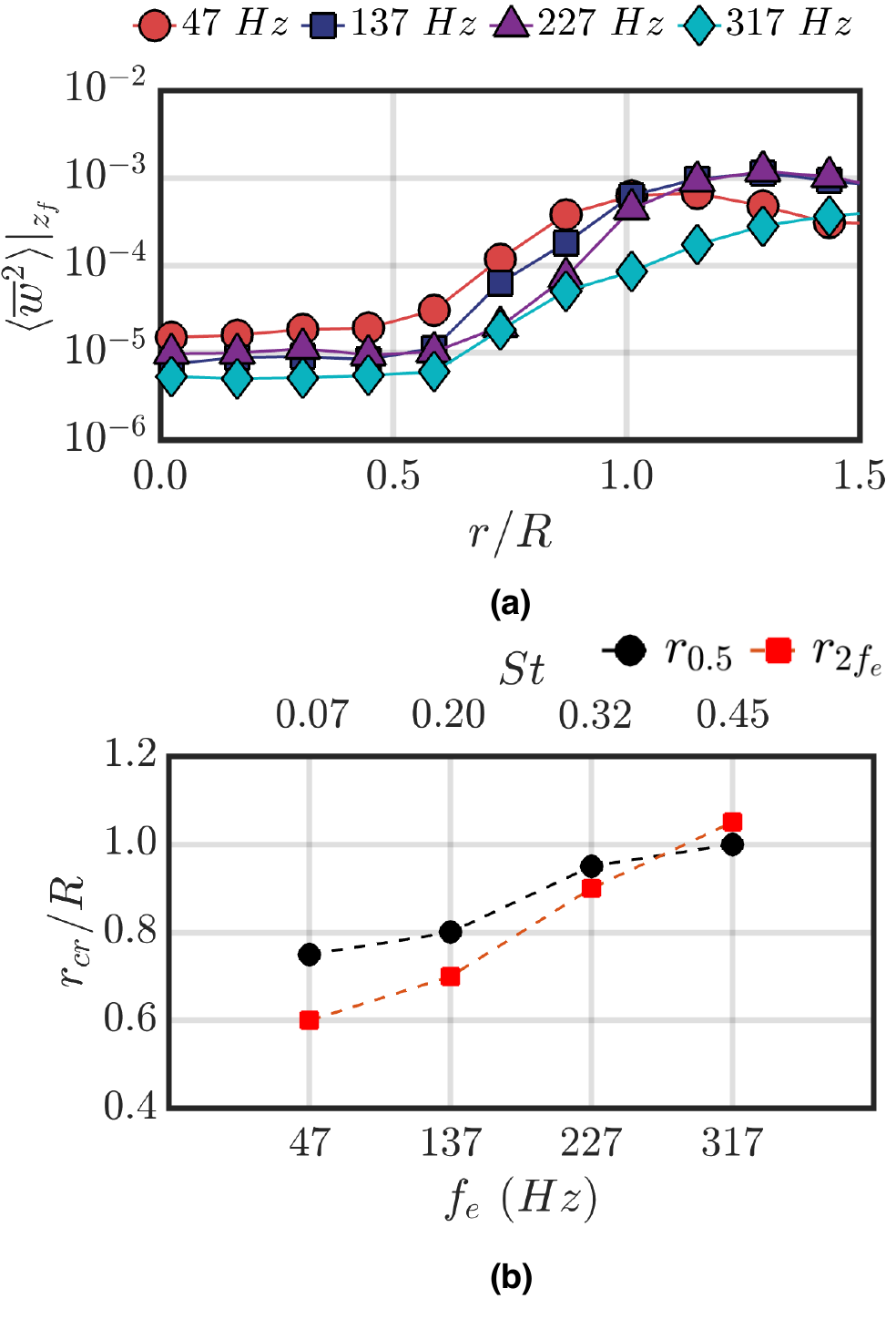}
    \caption[]%
    {{\small (a) Mean of squared normalized vorticity, $\langle \overline \omega^2 \rangle$, for different excitation frequencies at mean flame location (markers are displayed every 6 data points for clarity). (b) Comparison of the normalized critical radius where $\langle \overline \omega^2 \rangle$ is 50\% of its maximum, with normalized radial location where $2f_e$ starts to increase in $\hat P_{z_f}$ as a function of excitation frequency (shown in primary x-axis), and corresponding Stokes number, $\mathit{St}=f_e t_F$ (shown in secondary x-axis). }}  
    \label{fig:w_rms_zf}  
\end{figure}

\noindent\textbf{Q2: Is there a critical radius where second harmonics become dominant? } 
Recognizing that the emergence of higher harmonics at the outer edges of the flow is linked to the counter-rotating vortices in the shear layer (regions of high $\langle \overline \omega^2  \rangle$), we expect that the strength of these harmonics is closely tied to the intensity of the vorticity. Figure \ref{fig:w_rms_zf}(a) shows the radial distribution of $\langle \overline \omega^2  \rangle$, measured at the mean flame location, for all four excitation frequencies. As expected, vorticity remains negligible at small $r/R$, but rises sharply at larger $r/R$, i.e., within the shear layer. To identify a critical transitional radius where vorticity becomes significant, we define $r_{0.5}$ as the radial location at which $\langle \overline \omega^2  \rangle$ reaches 50\% of its maximum value. Figure \ref{fig:w_rms_zf}(b) compares $r_{0.5}$ across the four excitation frequencies with the corresponding radial positions where the second harmonic ($f/f_e=2$) in flame oscillations begins to grow (see Fig. \ref{fig:FFT_zf_harm}d). The results confirm that the transition points of $\langle \overline \omega^2  \rangle$ and the onset of second-harmonic growth in flame oscillations exhibit a consistent trend. It should be noted that alternative definitions of the transition radius (e.g., based on the maximum gradient or on 10\% of the maximum value) also yield qualitatively similar trends.

\textbf{Q3: Why do some excitation frequencies induce multiple higher harmonics, and some do not? } 
The combined effects of the imposed oscillation in the main flow and vortex shedding in the shear layer produce perturbations at both the excitation frequency ($f_e$) and its second harmonic ($2f_e$) near the flame’s outer edges. From nonlinear system dynamics, it is well known that when a system is perturbed by two frequencies (e.g., $f_1$ and $f_2$), the response can occur not only at these frequencies but also at their linear combinations (e.g., $f_1+f_2$, $2f_1+f_2$, $f_1+2f_2$, etc.) \citep{sujith2021dynamical}. Due to the counter-rotating vortex pair, the \textit{effective} excitation frequencies at the flame edges are $f_e$ and $2f_e$, which naturally lead to responses at $f_e$, $2f_e$, $3f_e$, etc. This trend is evident for $f_e=47$ Hz, where clear peaks are observed at all four frequencies in PSDs (Fig. \ref{fig:FFT_zf_harm}b and c for $\hat{P}_{z_f}$, and Fig. \ref{fig:psd_Kt_harm}b and c for $\hat{P}_{K_t}$). However, as the excitation frequency ($f_e$) increases, the Stokes numbers ($\mathit{St}_m = mf_e t_f$ for $m^{th}$ harmonics) of the higher harmonics also increase, diminishing the flame’s sensitivity to them. Consequently, for $f_e=137$ Hz, peaks are observed at $f_e$, $2f_e$, and $3f_e$; for $f_e=227$ Hz, only $f_e$ and $2f_e$ remain significant; and for $f_e=317$ Hz, the response is limited to $f_e$ alone.

\section{Summary}

In this study, we investigated premixed counterflow laminar flames subjected to oscillatory strain rates, with particular emphasis on the dynamics at off-center locations. The excitation frequency was systematically varied, with the oscillation amplitude prescribed using both the \textit{constant amplitude} and \textit{constant power} approaches. Flame dynamics were characterized using a combination of Mie-scattering imaging and Particle Image Velocimetry. The principal findings are summarized as follows:

\begin{itemize}
    \item Along the centerline, flame oscillations weaken with increasing excitation frequency. With an increase in frequency, the oscillation timescale decreases relative to the intrinsic flame timescale, and hence, the flame response diminishes accordingly.

    \item Spectral analyses of flame oscillations and flame-conditioned tangential strain rates show that near the counterflow core, the response is dominated by the imposed excitation frequency. Flow field analyses confirm that the core remains irrotational, such that the dynamics are primarily governed by the induced oscillatory excitation. 
    
    \item At the outer edge, however, the response is dominated by the second harmonic of the excitation frequency, along with additional higher harmonics. Flow analyses indicate that the imposed oscillation triggers vortex shedding within the shear layer, located at the outer edge of the flow. Each oscillation cycle produces a pair of counter-rotating vortices, which in turn drive flame oscillations at twice the excitation frequency.
    
    \item The shear-layer thickness and position determine the radial location at which the transition between excitation frequency-dominated vs. second harmonic-dominated dynamics occurs.
    
\end{itemize}







\section*{Acknowledgments}

The authors would like to thank Dr. Sombuddha Bagchi and Dr. Yue Weng for stimulating discussions and assisting with Matlab processing codes. The research was supported by the Office of Naval Research through the Naval Innovation, Science, and Engineering Center (NISEC) at UC San Diego (grant number:  N000142312831).

\section*{Supplementary material}

Supplementary Material includes additional datasets and figures supporting the main text: extended results for the \textit{constant amplitude} method, extended results for the \textit{constant power} method, reactive flame data for comparison with corresponding non-reactive cases, and bottom flame measurements for comparison with top flame data.

\bibliographystyle{elsarticle-num}
\bibliography{cnf-refs}
\end{document}